\documentclass[prl,twocolumn,superscriptaddress,floatfix,showpacs]{revtex4}
\usepackage{graphicx,amsfonts,amssymb,amsmath, hyperref}

\newif\ifhyper
\hypertrue
\ifhyper
\hypersetup{
   citecolor = {green},
   colorlinks = {true}, % false
   urlcolor = {blue} % magenta
}
\fi

\newcommand{\beq}{\begin{equation}}
\newcommand{\eeq}{\end{equation}}
\newcommand{\beqa}{\begin{eqnarray}}
\newcommand{\eeqa}{\end{eqnarray}}
\newcommand{\ket} [1] {\vert #1 \rangle}
\newcommand{\bra} [1] {\langle #1 \vert}
\newcommand{\braket}[2]{\langle #1 | #2 \rangle}

\newcommand{\tr}{\mathop{\mathrm{tr}}}

\def\bra#1{\langle#1\vert}
\def\ket#1{\vert#1\rangle}

\def\ipr#1#2{\langle#1\vert#2\rangle}

\def\Longarrow{\protect\@lra}
\def\@lra{\relbar\joinrel\relbar\joinrel\relbar\joinrel%
          \relbar\joinrel\rightarrow}

\begin{document}

\title{Topological transitions from multipartite entanglement with tensor networks: \\ a procedure for sharper and faster characterization }

\author{Rom\'an Or\'us}
\affiliation{Institute of Physics, Johannes Gutenberg University, 55099 Mainz, Germany}

\author{Tzu-Chieh Wei}
 \affiliation{C. N. Yang Institute for Theoretical Physics, State
University of New York at Stony Brook, NY 11794-3840, USA}

\author{Oliver Buerschaper}
\affiliation{Perimeter Institute for Theoretical Physics, 31 Caroline Street North, Waterloo, Ontario, Canada, N2L\,2Y5}
\affiliation{Dahlem Center for Complex Quantum Systems, Freie Universit\"at Berlin, 14195 Berlin, Germany}

\author{Artur Garc\'ia-Saez}
 \affiliation{C. N. Yang Institute for Theoretical Physics, State University of New York at Stony Brook, NY 11794-3840, USA}

\begin{abstract}
Topological order in a 2d quantum matter can be determined by the topological contribution to the entanglement R\'enyi entropies. However, when close to a quantum phase transition, its calculation becomes cumbersome. Here we show how topological phase transitions in 2d systems can be much better assessed by multipartite entanglement, as measured by the topological geometric entanglement of blocks. Specifically, we present an efficient tensor network algorithm based on Projected Entangled Pair States to compute this quantity for a torus partitioned into cylinders, and then use this method to find sharp evidence of topological phase transitions in 2d systems with a string-tension perturbation. When compared to tensor network methods for R\'enyi entropies, our approach produces almost perfect accuracies close to criticality and, on top, is orders of magnitude faster. The method can be adapted to deal with any topological state of the system, including minimally entangled ground states. It also allows to extract the critical exponent of the correlation length, and shows that there is no continuous entanglement-loss along renormalization group flows in topological phases.

\end{abstract}

\maketitle

\begin{figure}
 \includegraphics[width=9cm]{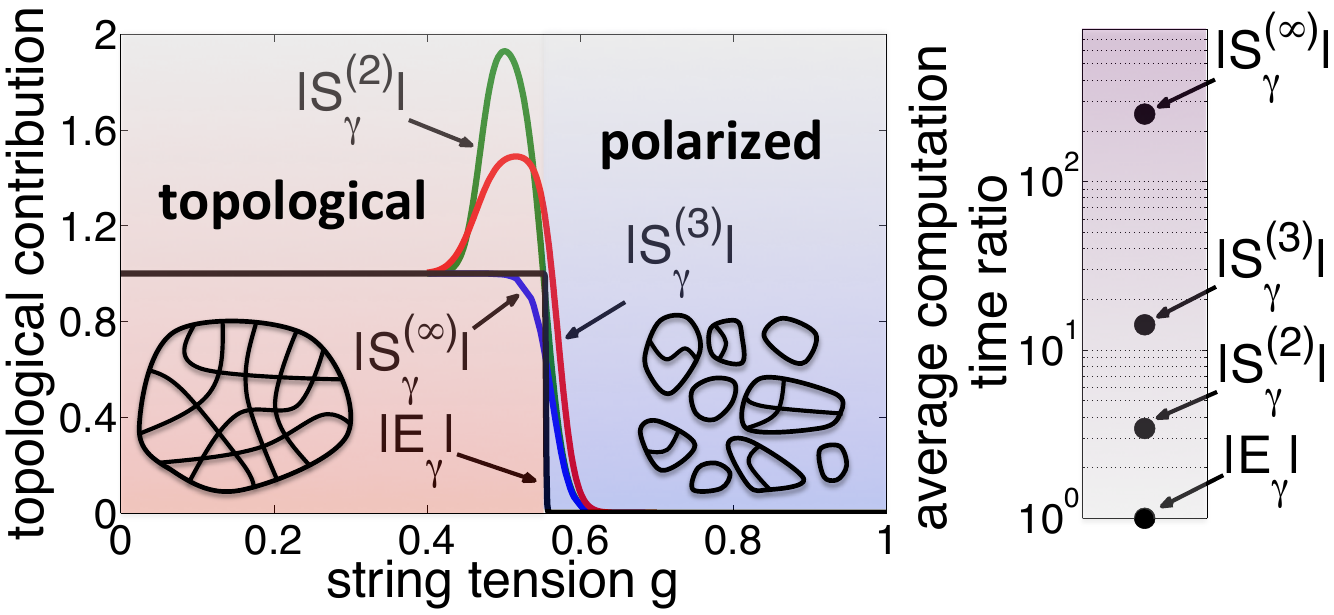}
  \caption{\label{Fig0}
  (color online) Left: phase transition for a topological 2d model with string tension. The black line corresponds to $|E_\gamma|$ as computed in Fig.\ref{Fig4}(a) for $n_b \rightarrow \infty$ using GE (to be explained later). The rest of the lines correspond to the best achievable calculation by the authors of the topological term $|S^{(n)}_\gamma|$ of the $n$th R\'enyi entropies $S^{(n)} = (1-n)^{-1}\log (\tr{(\rho^n)})$ of half an infinite cylinder for $n=2,3$ and $n=\infty$ (i.e. the single-copy entanglement \cite{sc}), using the methods  explained in the supplementary material. Compare also to similar calculations with tensor networks in, e.g., Ref.\cite{cost}.
  Typical sizes of string-nets populating the ground state for each phase are also represented. Right: average computation time ratio with respect to $|E_\gamma|$, for the different topological contributions.}
\end{figure}

Topological order \cite{topo} is a striking property of quantum matter beyond the Landau
paradigm and is characterized by an underlying pattern of
long-range entanglement. The existence of such a pattern can be detected, quantitatively, by the so-called
topological entanglement entropy $S_{\gamma}$ \cite{topoent}. Other entanglement properties are sensitive to topological
order as well \cite{topother}. Moreover, under the effect of a local
perturbation it is also well-known that topological order is
generally robust \cite{toric, robust} and can sustain a finite
perturbation. Intuitively, large closed strings and string-nets become
energetically expensive in a topological phase as a
string-tension is increased, thus ultimately favoring a transition
towards a topologically-trivial phase. A drawback of using entanglement to detect such topological transitions, however, is that it is very difficult to produce sharp numerical evidence. The reason for this is that commonly used methods, such as the calculation of the topological contribution in R\'enyi entropies \cite{Renyi}, suffer from a significant drop in accuracy when close to a quantum critical point \cite{cost}, see Fig.\ref{Fig0}. Here we show how multipartite entanglement, in combination with tensor networks, improves accuracies to an almost perfect level and, on top, is computed orders of magnitude faster than any R\'enyi entropy.

More specifically, here we use a \emph{novel and efficient tensor
network method to evaluate the topological contribution to the
geometric entanglement (GE) of blocks,  which we call $E_{\gamma}$,
for a torus partitioned into cylinders}. When close to a quantum
phase transition, we find that this approach completely outperforms
in accuracy and efficiency calculations of R\'enyi entropies on
infinite cylinders with tensor networks \footnote{See, e.g.,
Ref.\cite{ig}, as well as the supplementary material for the
specific algorithms implemented to obtain the data in
Fig.\ref{Fig0}.}. Without describing the technical details, the main
result is summarized in Fig.\ref{Fig0}. We apply a string
tension $g$ (which corresponds to a magnetic field in the Hamiltonian~\cite{mappingIsing,magneticField}) to certain toric code ground states, and compute the
topological contribution of the GE of these ``strained" toric code
states. Unlike the topological R\'enyi entropies, the computed
$E_\gamma$ stays close to $-1$ throughout the entire topological
phase and, as the string tension increases, it sharply drops down to
zero at $g^*\approx 0.56$ and remains there in the trivial phase.
From a mapping to a classical 2D Ising model~\cite{mappingIsing} we
obtain analytically a transition at $g^*=\sqrt{1+\sqrt{2}}-1\approx 0.5537$, in
agreement with the above. $E_\gamma$ is thus an excellent tool to
pinpoint topological phase transitions.

Our method uses Projected Entangled Pair States (PEPS) \cite{PEPS}.
If the PEPS is topological \cite{topoPEPS}, then a robust
$E_{\gamma}$ is extracted via finite-size scaling for large
toruses. For non-trivial partitions we find $E_\gamma$ to
depend on the particular superposition of ground states on the
torus, in agreement with the behavior of R\'enyi entropies and entanglement entropy
 \cite{mes}. Our calculations focus mainly on PEPS with a translation invariant representation. This has two main advantages: first, it simplifies the calculations, and second, it corresponds to the type of unique ground state of a topological system that can be found on an infinite plane using, e.g., the iPEPS method \cite{iPEPS}. Even if such states have a weaker topological contribution on a torus than minimally entangled states (MES), they are much simpler to deal with, and already produce non-trivial topological contributions. In any case, we shall see that our method can be easily extended to PEPS representations that are not invariant under translations, thus including MES if necessary. Importantly, with this method we also have access to other properties. For instance, in the supplementary material we show how to extract the critical exponent $\nu$, and how to see that there is no continuous entanglement-loss along renormalization group flows in topological phases  \cite{loss} together with a fidelity analysis \cite{fid}.

 \begin{figure}
 \includegraphics[width=7.5cm]{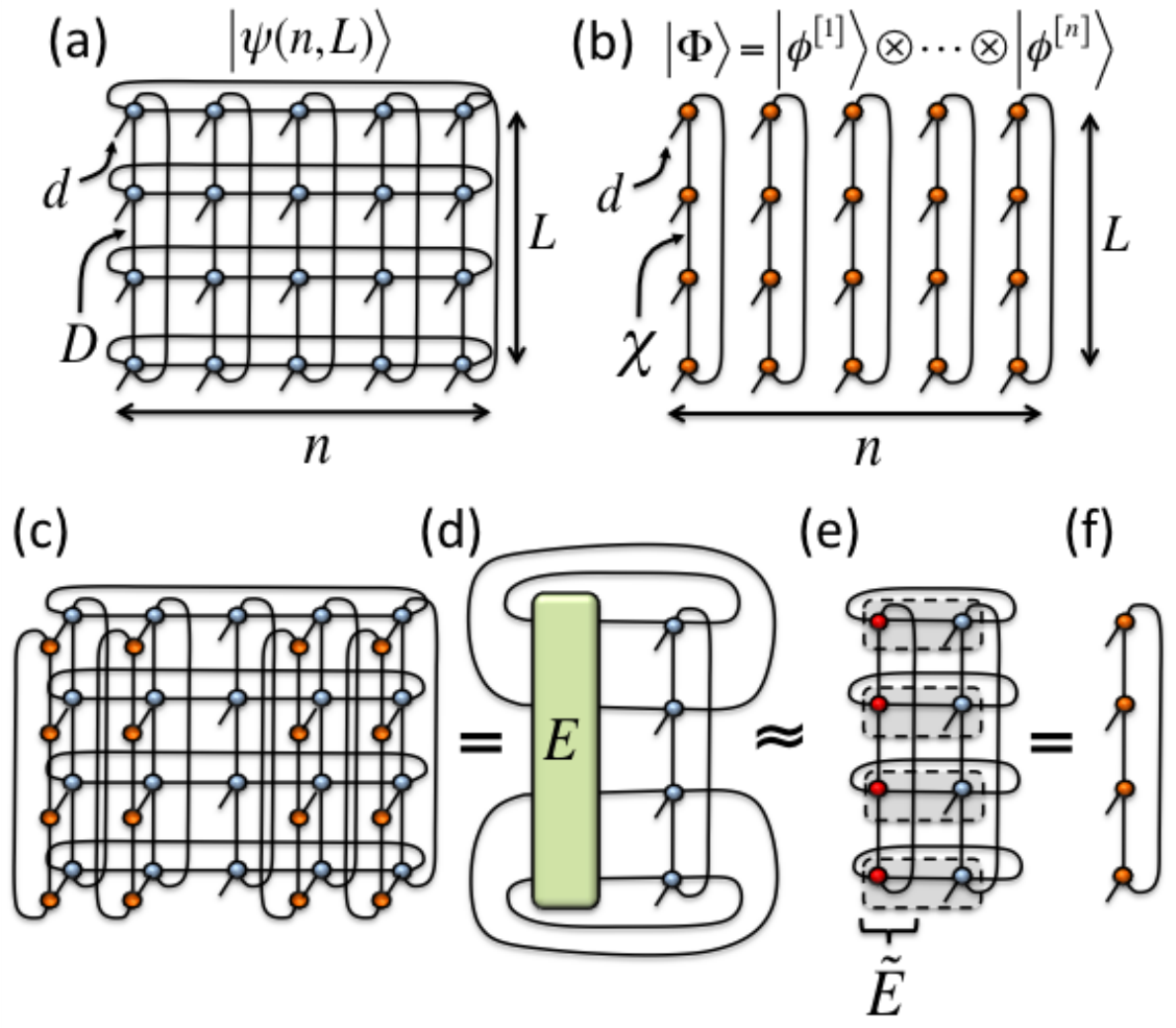}
  \caption{\label{Fig1}
  (color online) (a) PEPS $\ket{\Psi(n,L)}$ wrapped around a torus. (b) Product state $\ket{\Phi}$ of MPSs with pbc (cylinders with $l=1$). (c) Contraction to compute the optimal state for a given cylinder. (d) Exact result of the contraction in (c) in terms of the environment tensor $E$. (e) Approximation of $E$ by an effective environment $\tilde{E}$ described by an MPO. (f) Resulting optimal MPS for the cylinder in an iteration step (see main text).}
\end{figure}

\emph{GE and topological GE ---} The geometric entanglement of
blocks \cite{ge, GEblocks} has recently proven useful to assess
topological order \cite{topoGE, inprep}. This multipartite measure  has been
extensively used in  quantum phase transitions \cite{GEqpt}, and can
be measured experimentally, e.g., in NMR \cite{exper} and potentially
in optical lattice experiments \cite{opticalLattice}. In contrast to
all other entanglement approaches for topological matter, the GE takes
into account the multipartite structure of entanglement in quantum
many-body states. It amounts to computing the closest product state
$\ket{\Phi}$ to a given quantum state $\ket{\Psi}$ in the Hilbert
space, where the product state has a separable structure of $n_b$
blocks, i.e., $\ket{\Phi}Ê= \prod_{i=1}^{n_b} \ket{\phi^{[i]}}$. It
thus quantifies the merit of a possible mean-field
description of the quantum state. Conveniently, the GE is defined as $E_G \equiv -\log |\braket{\Phi}{\Psi}|^2$.

One of the latest findings has been
that, for renormalization group (RG) fixed points such as the toric
code and other topological exactly-solvable models, the GE of blocks obeys $E_G = E_0 - E_{\gamma}$, with
$E_{\gamma}$ a topological contribution (the topological GE) and
$E_0$ some non-universal term \cite{topoGE}. It was observed that $E_{\gamma} =
S_{\gamma}$ for the considered models. This constant contribution was shown to be directly connected to the size of the gauge group, which in turn governs topological order in the system. As for $E_0$ it was found that $E_0 \propto n_b L$, with $n_b$
the number of blocks with a contractible boundary
of size $L$. Moreover, under perturbations it was argued
that $E_G = E_0 - E_{\gamma} + O(L^{-\nu'})$ for $L \gg 1$,  where again $E_0 \propto n_b L$, $\nu
'$ is some exponent, and $E_{\gamma}$ is the (robust) topological
term. \color{black}ÊRecently, the topological GE has also been used to identify minimally-entangled ground states, both for abelian and non-abelian models \cite{inprep}.\color{black}

\emph{Computing $E_G$ and $E_\gamma$ from a PEPS ---} Our approach
to computing the GE of non-contractible blocks $E_G$ for  large block
sizes and its topological contribution $E_\gamma$ uses 2d PEPS and
1d Matrix Product States (MPS). Both PEPS and MPS have been widely
discussed in the literature (see, e.g., Ref.\cite{tn}). It is worth mentioning that PEPS can describe 2d topological
phases naturally, both chiral \cite{chiralPEPS} and non-chiral (or
doubled) \cite{topoPEPS}. For simplicity, here we focus on
non-chiral topological order, but a generalization of our method to
chiral models is also possible.

As a starting point we assume that a PEPS $\ket{\Psi}$ with
(potentially) topological order is given on a torus of $n
\times L$ sites, see Fig.\ref{Fig1}(a). We call such a state
$\ket{\Psi(n,L)}$. This PEPS could be the result of an analytical
derivation, or have been computed numerically from a Hamiltonian
using, e.g., the iPEPS algorithm  \cite{iPEPS} and wrapping later its
tensors around a finite torus.

The goal now is to extract $E_G$ and $E_\gamma$ from such a PEPS.
With this in mind, we partition the torus into $n_b$ \emph{cylinders} of equal width $l=n/n_b$. Thus each cylinder contains $l \times L$ sites, see, e.g., Fig.\ref{Fig1}(b). This choice of
partition will have a double benefit. First, it will simplify the
tensor contractions in the method. Second, it will be sensitive to different ground states on the torus, and hence to MES.

We focus on the case $l=1$, so that $n_b = n$. In this case, one needs to find the
closest product state $\ket{\Phi} = \otimes_{i=1}^{n_b}Ê\ket{\phi^{
[ i ]}}$ of cylinders of one-site width to $\ket{\Psi(n,L)}$, with
$\ket{\phi^{ [ i ]}}$ the state for cylinder $i$. To do such a calculation
efficiently, we further approximate $\ket{\phi^{ [ i ]}}$ for each
cylinder by an MPS of $L$ sites with periodic boundary conditions (pbc)
and bond dimension $\chi$, see Fig.\ref{Fig1}(b) \footnote{This is indeed justified, as we show in the supplementary material.}. Thus, \emph{the
original problem is reduced to finding the product state of MPSs
with pbc that maximizes the overlap with a given PEPS on a torus},
which is a well-posed tensor network problem.

In what follows we describe an optimization procedure, well-suited
for gapped  topological phases, to solve this problem. The method
assumes a translation invariant PEPS, but it can also be generalized
to PEPS without translation symmetry (such as MESs).

\emph{1.- Assume translation invariance so that cylinders are repeated periodically}. While not
necessary for a finite system, this assumption simplifies the
calculations and also produces good results for
translation invariant PEPS. Here a 2-cylinder unit-cell is already
sufficient, but bigger unit cells can also be considered.

\emph{2- Fix all tensors in the MPSs to some initial (e.g., random)
values except for one cylinder, and optimize variationally the MPS
tensors for that cylinder}. The result of this optimization is given by
the diagram in Fig.\ref{Fig1}(c,d). Notice, though, that for a 2d
lattice the environment tensor $E$ cannot be computed both exactly
and efficiently, and therefore needs to be approximated.

\emph{3.- Compute an effective environment $\tilde{E}$ approximating
the exact environment $E$},  using some method to approximate
contractions of 2d tensor networks. In our case we assume further translation invariance within each cylinder, and use the iTEBD
method for non-unitary evolutions \cite{itebd1,itebd2}, without explicitly implementing the boundary conditions imposed by the torus
geometry, and adapted to deal with Matrix Product Operators (MPO).
The specifics are explained in the supplementary material. As a result of this approach,
an infinite MPO of bond dimension $\chi'$ is produced which is then cut at length $L$ and
wrapped around a circle with pbc. Such an approximation is particularly accurate for large $L$ and gapped phases, which is precisely the regime of interest to extract $E_\gamma$. This finite MPO with pbc describes
the effective environment $\tilde{E}$, see Fig.\ref{Fig1}(e).

 \emph{4.- Approximate the optimal MPS for the cylinder} as in Fig.\ref{Fig1}(f).

 \emph{5.- Substitute this MPS in all the equivalent cylinders} by translation invariance.

 \emph{6.- Repeat the procedure} for the next cylinder in the unit cell.

 \emph{7.- Iterate} until convergence.

The optimal overlap is thus evaluated as
\beq
\Lambda_{\max}(n,L) \simeq \frac{|\ipr{\Phi}{\Psi(n,L)}|}{\sqrt{|\ipr{\Phi}{\Phi}| |\ipr{\Psi(n,L)}{\Psi(n,L)}|}},
\eeq
with $\ket{\Phi} = \otimes_{i=1}^{n_b}Ê\ket{\phi^{ [ i ]}}$, and $\ket{\phi^{ [ i ]}}$ the optimal MPS for each cylinder. In this expression, the numerator can be approximated  using, e.g., the procedure described in the first section of the supplementary material with a computational cost of $O(n_b \chi^3 \chi'^3 D^5 + L\chi'^3)$. The norm of $\ket{\Phi}$ is simply the product of the norms of the $n_b$ MPS of size $L$ with pbc, which can be evaluated exactly and efficiently in $O(L \chi^5)$ steps (see, e.g., \cite{tn}). The norm of the $n \times L$ PEPS $\ket{\Psi(n,L)}$ can be approximated as in the first section of the supplementary material with a computational cost of $O(n (\chi''^2 D^9 + \chi''^3 D^6) + L\chi''^3)$, with $\chi''$ the bond dimension of the needed MPO. Finally, the GE is given by  $E_G(n,L)\equiv-\log_2\Lambda^2_{\max}(n,L)$.

For an $n \times L$ PEPS on a torus it is thus possible to
approximate $E_G(n,L)$ as above. To get the
topological contribution, the next step is to perform
\emph{finite-size scaling} with respect to $n$ and $L$. In
particular, we have $E_G(n \gg 1, L \gg 1) \sim \alpha n L -
E_\gamma(n,L)$, where $E_\gamma(n,L)$ includes both the topological
component $E_\gamma$ as well as finite-size corrections. We can then fix $n$ and compute $E_G(n, L)$ for increasing $L$. Doing a
linear fit for large $L$ allows us to extract an approximation to
the topological GE by extrapolating the fit down to $L=0$. The larger $n$ is, the more accurate the
approximation is. Thus, the value of the topological correction is
finally estimated as $E_\gamma = \lim_{n,L \rightarrow \infty}
E_\gamma(n,L)$.

\begin{figure}
 \includegraphics[width=9.1cm]{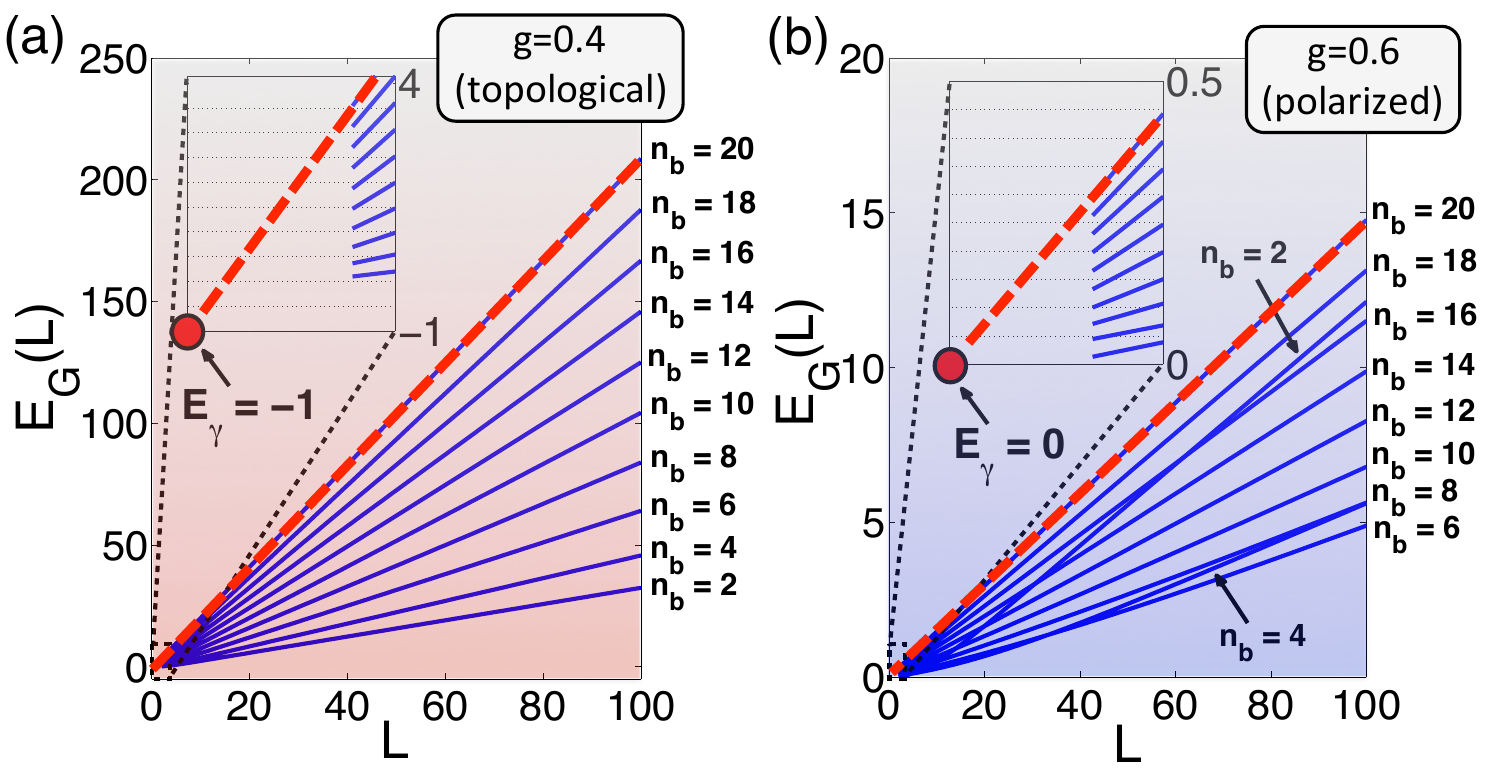}
  \caption{\label{Fig3}
  (color online) $E_G(n,L)$ for the toric code with string tension on the square lattice (similar results are also obtained for the honeycomb lattice). (a) Case $g=0.4$. The extrapolation of the linear fit (red dashed line) for, e.g., $n=20$ hits $L=0$ around $\sim -1$ (red dot), as expected in the topological phase. (b) Case $g=0.6$. The same calculation yields $\sim 0$, as expected in the polarized phase.}
\end{figure}

Some remarks are in order. First, accuracy can always be improved by increasing
the different bond dimensions, or by applying tensor network methods that explicitly take into
account pbc rather than iTEBD, or by using larger unit
cells (and even breaking completely translation invariance along any direction) in the
product state $\ket{\Phi}$. Second, cylinders of larger width $l > 1$ can be considered by using a PEPS for an
$l$-leg ladder with pbc to approximate the state $\ket{\phi^{[i]}}$
within each cylinder, or perhaps even an MPS with pbc and physical
dimension $d^l$ (with $d$ the physical dimension of a single site). An example of such a calculation is shown in the supplementary material. Third, MESs can also be studied introducing
minor changes in the method. For this, notice that the PEPS representation of an MES
is translation invariant except for, e.g., one cylinder where a Wilson loop operator acted. Thus, one chooses
$\ket{\Phi}$ as a translation invariant product state of MPSs,
except for the Wilson loop cylinder, where a different MPS is chosen.
The rest of the method just follows. \color{black} Fourth, the method relies on single-layer contractions of a $2d$ tensor network, which are computed both more efficiently and more accurately than the double-layer contractions in R\'enyi entropy calculations. This is model-independent, and explains the overall computational gain from Fig.\ref{Fig0}. A more detailed justification is provided in the supplementary material. \color{black}

\begin{figure}
 \includegraphics[width=8cm]{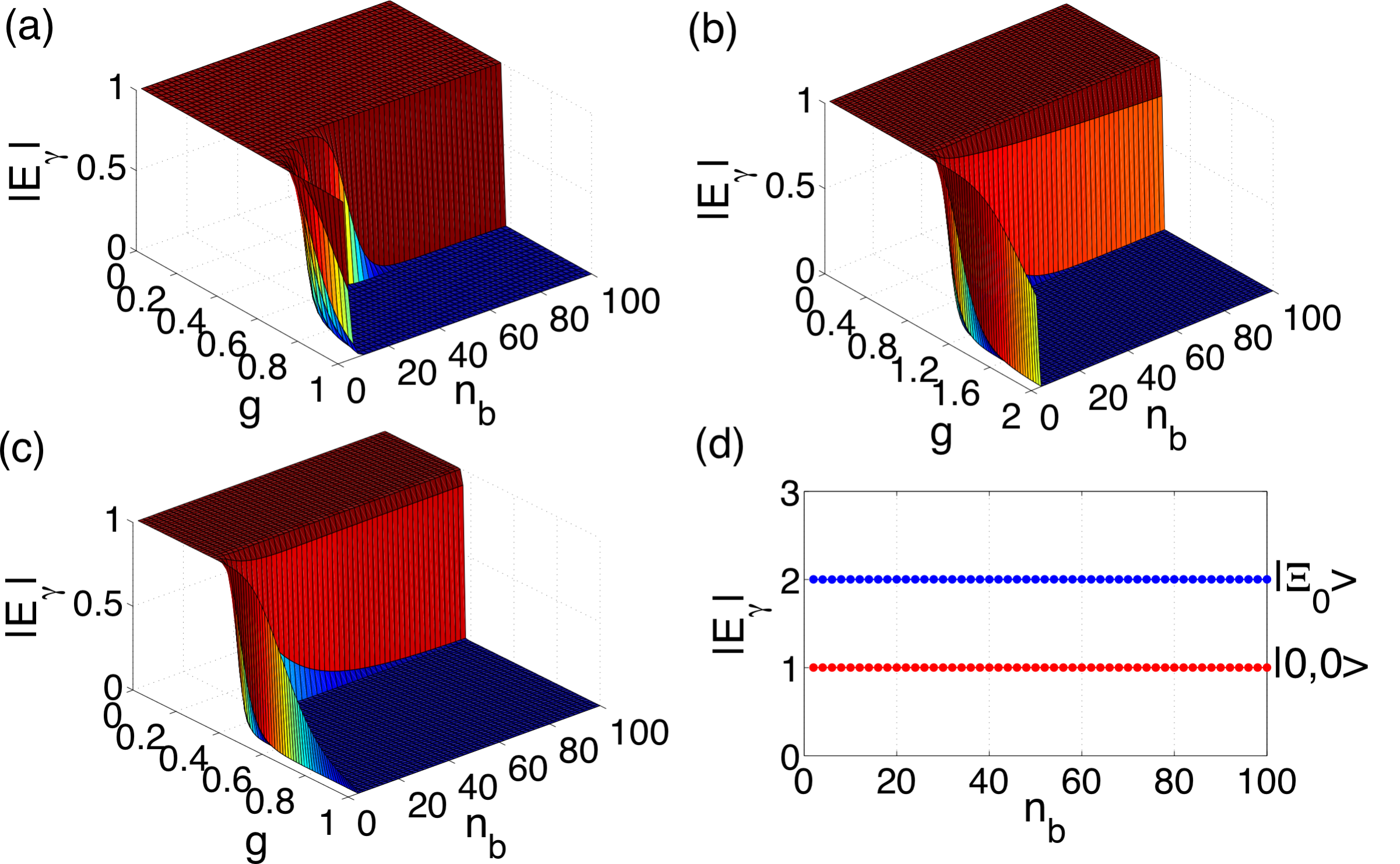}
  \caption{\label{Fig4}
  (color online) absolute value of $E_\gamma$ extrapolated from the scaling with $L$ as in
  Fig.(\ref{Fig3}), as a function of the string tension $g$ (for a-c) and the number of blocks
  $n_b$ (for a-d). Plots are for the perturbed (a) $\ket{0,0}$ state on the square lattice,
  (b) $\ket{0,0}$ state on the honeycomb lattice, and (c) $\ket{+,+}$ state on the honeycomb
  lattice. Notice that (b) and (c), though being ground states on the same lattice,
  have different transition points. This is because the string tension~$g$ was
  applied in different bases, hence corresponds to different physical perturbations.
  However, we have also checked that, when the
\emph{same} perturbation is applied to different ground states on
the same lattice, the topological phase transition takes place at
the same critical point, showing that the transitions in $E_\gamma$
do \emph{not} depend on the specific choice of ground state. Figure (d)
corresponds to the unperturbed toric code on a square lattice for
two different ground states: an MES $\ket{\Xi_0}$, and a non-MES
$\ket{0,0}$.}
\end{figure}

\emph{Topological phase transitions from $E_{\gamma}$ ---} Using the above method we
computed $E_G$ and $E_\gamma$ for the toric code model \cite{toric}
with string tension on the square and honeycomb lattices. Details
about the PEPS for these models \color{black} as well as about the blocking schemes \color{black} are given in the supplementary
material. The string tension $g$ drives the systems towards a phase
transition between topological and polarized phases. Using the
notation from Ref.\cite{topoGE},  we considered perturbations to two
non-equivalent ground states $\ket{0,0}$ and $\ket{+,+}$ for the
honeycomb lattice, whereas for the square lattice we considered
perturbations to the $\ket{0,0}$ state. In the topological phase, these states are the unique ground states
of the system on an infinite plane, but on a torus they correspond
to a superposition of MESs with topological entropy $S_\gamma = -1$
for a non-contractible bipartition \cite{mes}.

Our calculations were done for toruses up to $n=100$ and $L=100$.  Larger
sizes could have easily been considered if necessary. In
Fig.\ref{Fig3} we show an example of the scalings with $L$ for
different values of $n$ up to $n=20$ for two different string
tensions $g=0.4,0.6$ on the square lattice. The linear fit is
computed from the last half of $L$ values, which  produces
robust results. In the plots, the extrapolation of the fit to $L=0$
hits the vertical axis around $-1$ if $g$ is small, corresponding to
the topological phase, and around $0$ for large $g$, corresponding
to the polarized phase. From the fits we can extract $E_\gamma$ as a
function of $n=n_b$ and $g$, as shown in Fig.\ref{Fig4}(a-c) for the
three states considered. Remarkably, these plots show very sharp
indications of topological phase transitions for all these models
for large $n_b$. With this approach we also extracted $E_\gamma$ for one of the MES
of the square lattice toric code on a torus without perturbation.
Specifically, we considered the state $\ket{\Xi_0} \equiv 2^{-1/2}
(\ket{0,0} + \ket{1,0})$ (in the notation of Ref.\cite{topoGE})
which has $S_\gamma = -2$ for a non-trivial torus bipartition
\cite{mes}. Remarkably, we also find $E_\gamma = -2$ for this state, see
Fig.\ref{Fig4}(d).

\emph{Conclusions---} We obtained sharp evidence of topological
quantum phase transitions for 2d systems, by calculating $E_\gamma$
using a new and efficient tensor network method for non-trivial
partitions on a torus. Our method completely outperforms similar tensor network calculations of R\'enyi entropies for infinite cylinders, by being orders of magnitude more accurate and efficient close to criticality \cite{cost}. This approach can also be applied to different ground states, including MES, and allows to extract other non-trivial information about the system (e.g., correlation length critical exponent, and lack of continuous entanglement loss along RG flows in topological phases). Our work motivates further research along several directions. For instance, it would be possible to use these tools to study chiral topological
order \cite{poll}, topological critical exponents, and MES. Beyond tensor network methods, it would be interesting to study how to compute $E_\gamma$ using Quantum Monte Carlo, and compare the accuracy and efficiency to that of  2-R\'enyi entropy calculations.

\acknowledgements Discussions with B. Bauer, F. Pollmann, A.
Sanpera and G. Vidal are acknowledged. T.-C.W. and A.G.-S. acknowledge the
support by the National Science Foundation under Grants No. PHY
1314748 and No. PHY 1333903.

\clearpage

\onecolumngrid
\section{Supplementary material}

In this supplementary material we provide details on the following:

\begin{enumerate}
\item Tensors for the toric code model with string tension on different lattices.
\item Calculation of the effective environment using iTEBD and MPOs.
\item Blocking of spins on the square and honeycomb lattices. 
\item Extracting $E_\gamma$ using cylinders of width $l=2$ on the square lattice.
\item Alternative optimization strategies and finite-size effects in $E_G$.
\item Fidelity diagram and no entanglement loss along RG flows in topological phases.
\item Extracting the critical exponent $\nu$ from the finite-size scaling of $E_\gamma$.
\item Efficient methods to compute the $n$th-R\'enyi entropy and its topological contribution, for finite and infinite $n$, for a 2d PEPS on an infinite cylinder.
\item Numerical truncation errors in $E_G$ and R\'enyi entropies.

\end{enumerate}

\subsection{1. Tensors for the toric code with a string tension}

As examples of perturbed topological models, in this paper we have considered Kitaev's toric code model \cite{atoric} with a string tension on square and honeycomb lattices. It is well known that the ground states for these systems on an infinite plane can always be specified by a PEPS. Here we adopt the convention that tensor bond indices are subscripts, starting from the leftmost index in the lattice and following a clockwise rotation order, and physical indices are superscripts. \color{black} Details on how to derive the PEPS tensors for the toric code can be found on, e.g., Ref.\cite{annals}. \color{black} According to this convention, and following the notation of Ref.\cite{atopoGE} for the toric code ground states, the three PEPS that we consider here are:

\vspace{5pt}
 \emph{(i) Perturbed $\ket{0,0}$ ground state on the square lattice}. This is described by $2$ PEPS tensors $A$ and $B$ with bond dimension $D=2$. The non-zero coefficients $A^i_{\alpha \beta \gamma \delta}$ (where $i$ and $\alpha \ldots \gamma$ are the physical and bond indices respectively) are given by
\begin{eqnarray}
A_{1,1,1,1}^1 = 1+g  &\hspace{10pt}& A_{2,2,1,1}^2 = 1 \nonumber \\
A_{2,2,2,2}^1 = 1+g &\hspace{10pt}&  A_{1,1,2,2}^2 = 1
\label{tcsq}
\end{eqnarray}
with $g$ a string tension, and $B$ being a rotation of $\pi/2$ of $A$ on the lattice.

\begin{figure}
 \includegraphics[width=9cm]{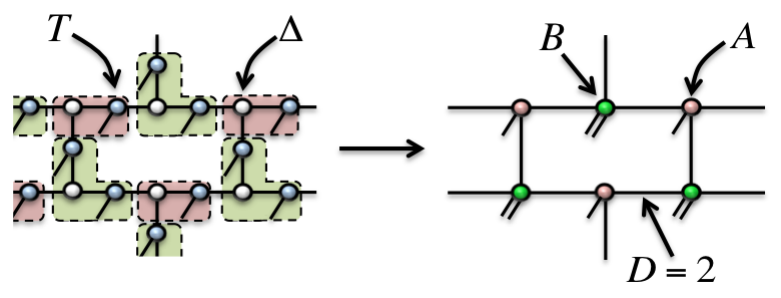}
  \caption{\label{FigAp1}
  (color online) tensors for the PEPS $\ket{0,0}$ perturbed ground state on the honeycomb lattice. The lattice is represented as a brickwall.}
\end{figure}

\vspace{5pt}
 \emph{(ii) Perturbed $\ket{0,0}$ ground state on the honeycomb lattice}. This is described by two tensors $T$ and $\Delta$ with non-zero coefficients
 \begin{eqnarray}
T_{1,1}^1 = 1+g  &\hspace{10pt}& T_{1,2}^2 = 1 \nonumber \\
T_{2,2}^1 = 1+g &\hspace{10pt}&  T_{2,1}^2 = 1
\end{eqnarray}
and
\beq
\Delta_{1,1,1} =  \Delta_{2,2,2} = 1 ,
\eeq
see Fig.\ref{FigAp1}. A more convenient description for the numerical calculations is given in terms of tensors $A$ and $B$ computed from $T$ and $\Delta$ as shown in Fig.\ref{FigAp1}. In this construction the PEPS bond dimension is $D=2$, whereas the physical dimensions are $2$ for $A$ and $4$ for $B$.

\begin{figure}
 \includegraphics[width=9cm]{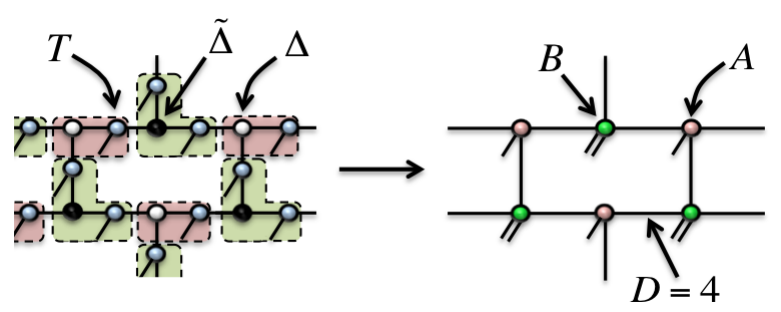}
  \caption{\label{FigAp2}
  (color online) tensors for the PEPS $\ket{+,+}$ perturbed ground state on the honeycomb lattice. The lattice is represented as a brickwall.}
\end{figure}

\vspace{5pt}
 \emph{(iii) Perturbed $\ket{+,+}$ ground state on the honeycomb lattice}. This is described by tensors $T, \Delta$ and $\widetilde{\Delta}$ with non-zero coefficients
  \begin{eqnarray}
T_{1,1}^+ = 1+g &\hspace{10pt}& T_{2,2}^- = 1\nonumber \\
T_{4,4}^+ = 1+g &\hspace{10pt}& T_{3,3}^- = 1
\end{eqnarray}
\begin{eqnarray}
\Delta_{1,1,1} = \Delta_{3,3,1} &=& \Delta_{2,1,3} = \Delta_{1,2,2} = 1 \nonumber \\
\Delta_{4,3,3} = \Delta_{3,4,2} &=& \Delta_{2,2,4} = \Delta_{4,4,4} = 1
\end{eqnarray}
\begin{eqnarray}
\widetilde{\Delta}_{1,1,1} = \widetilde{\Delta}_{3,3,1} &=& \widetilde{\Delta}_{1,2,3} = \widetilde{\Delta}_{2,1,2} = 1 \nonumber \\
\widetilde{\Delta}_{3,4,3} = \widetilde{\Delta}_{4,3,2} &=& \widetilde{\Delta}_{2,2,4} = \widetilde{\Delta}_{4,4,4} = 1 \ ,
\end{eqnarray}
see Fig.\ref{FigAp2}. One can rewrite again the tensor network in terms of two tensors $A$ and $B$ as shown in Fig.\ref{FigAp2}. This time, the PEPS bond dimension is $D=4$, and the physical dimensions are again $2$ for $A$ and $4$ for $B$.

\subsubsection{Remarks}
The above perturbations on the $\ket{0,0}$ and $\ket{+,+}$ states on the honeycomb lattice correspond, in fact, to different physical perturbations: one adds weight to strings of $1$'s or $+$'s respectively. It is, however, easy to apply the same perturbation to both ground states. For instance, in the $\{ \ket{1}, \ket{2} \}$ basis for every spin, the perturbed $\ket{0,0}$ state can be written as $Q^{\otimes N} \ket{0,0}$, with $N$ the number of spins and $Q = (1+g)\ket{1}\bra{1} + \ket{2}\bra{2}$. The perturbed $\ket{+,+}$ state after applying the \emph{same} perturbation $Q$ to every spin then reads $Q^{\otimes N} \ket{+,+}$, and the PEPS for such a state is easy to compute in the $\{ \ket{+}, \ket{-} \}$ local basis (with $\ket \pm = (\ket{1} \pm \ket{2})/\sqrt{2})$) and noticing that the perturbation operator can be written as $Q = (1+g/2)(\ket{+}\bra{+} + \ket{-}\bra{-}) +(g/2)(\ket{+}\bra{-} + \ket{-}\bra{+})$

\subsection{2. Computation of the effective environment $\tilde{E}$ as an MPO using iTEBD}

\begin{figure}[h]
 \includegraphics[width=10cm]{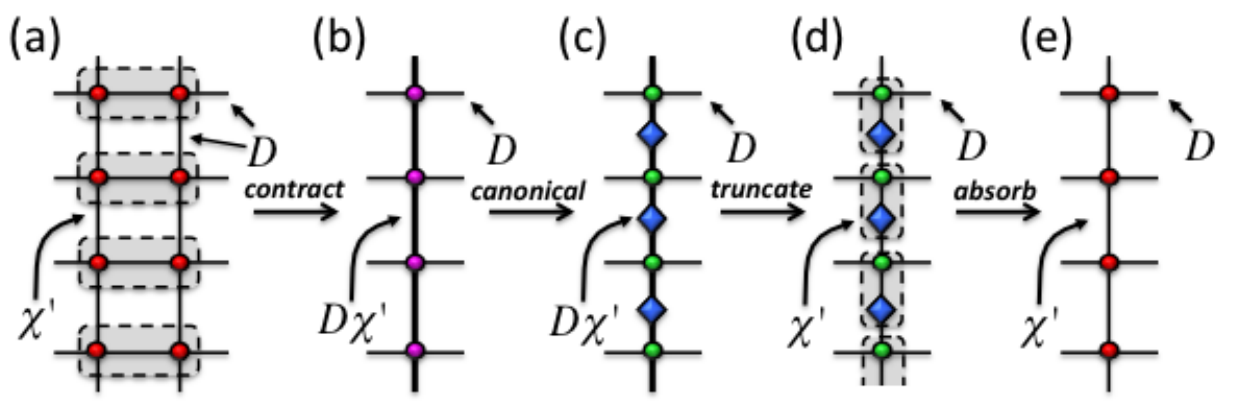}
  \caption{\label{Fig2}
  (color online) iTEBD method for non-unitary MPOs, adapted from Ref.\cite{aitebd2}. (a) Contraction of 2 MPOs of bond dimensions $\chi'$ and $D$. (b) Resulting MPO of bond dimension $D \chi'$. (c) Canonical form of the MPO in (b), in terms of MPO tensors (circles) and matrices of Schmidt coefficients (diamonds). (d) Truncated MPO of bond dimension $\chi'$ in canonical form. (e) Final MPO, were we have absorbed the Schmidt coefficients into the tensors at each site.}
\end{figure}

At several points in our numerical methods we use iTEBD to produce an MPO approximation of the multiplication of several transfer operators, e.g., when computing the exact environment $E$ for $E_G$ in step 3. To do so, we assume that we have translation invariance every, e.g., 2 lattice sites at least. This MPO is produced is as follows:

\emph{1.- Contract} the tensors for two columns as in
Fig.\ref{Fig2}(a). This is the product of two infinite MPOs. If both
MPOs have bond dimension $D$, then the resulting MPO will have bond
dimension $D^2$.

\emph{2.- Find the canonical form} of the resulting
MPO, see Fig.\ref{Fig2}(b,c). This can be done using the algorithms
from Ref.\cite{aitebd1, aitebd2}, adapted to an MPS with two physical legs per
site of dimension $D$ each.

\emph{3.- Truncate}  the bond dimension of the MPO to
its largest $\chi'$ Schmidt coefficients of a bipartition across
each link, see Fig.\ref{Fig2}(d).

\emph{4.- Absorb}  the Schmidt coefficients into the
tensors at each site, see Fig.\ref{Fig2}(e).

\emph{5.- Iterate} until approximating the
contraction of $n_b-1$ infinite columns. At each iteration step we
will have as input two infinite MPOs, one of bond dimension $\chi'$
and another one of bond dimension $D$. The output will always be a new
infinite MPO of bond dimension $\chi'$.

%%%%%%%%

\color{black}
\subsection{3. Choice of blocking}

\begin{figure}[h]
 \includegraphics[width=9cm]{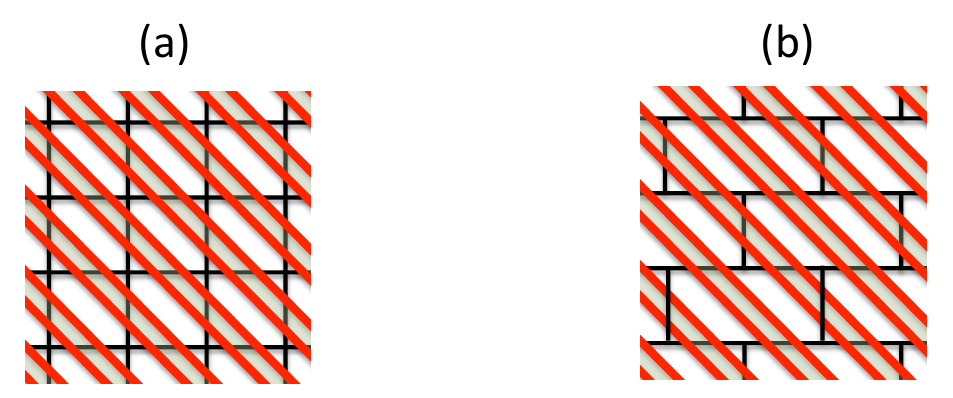}
  \caption{\label{figextra}
  (color online) Blocking of spins for (a) square lattice, and (b) honeycomb lattice. Spins are on the links.}
\end{figure}

In Fig.\ref{figextra} we show our explicit choice of blocks for the square and honeycomb lattices, for the case of blocks of width $l=1$. Larger widths ca be considered easily following this scheme. In the lattices, spins are on the links. 
\color{black}

%%%%%%%%%%

\subsection{4. Extracting $E_\gamma$ using cylinders of width $l=2$}

\begin{figure}[h]
 \includegraphics[width=11cm]{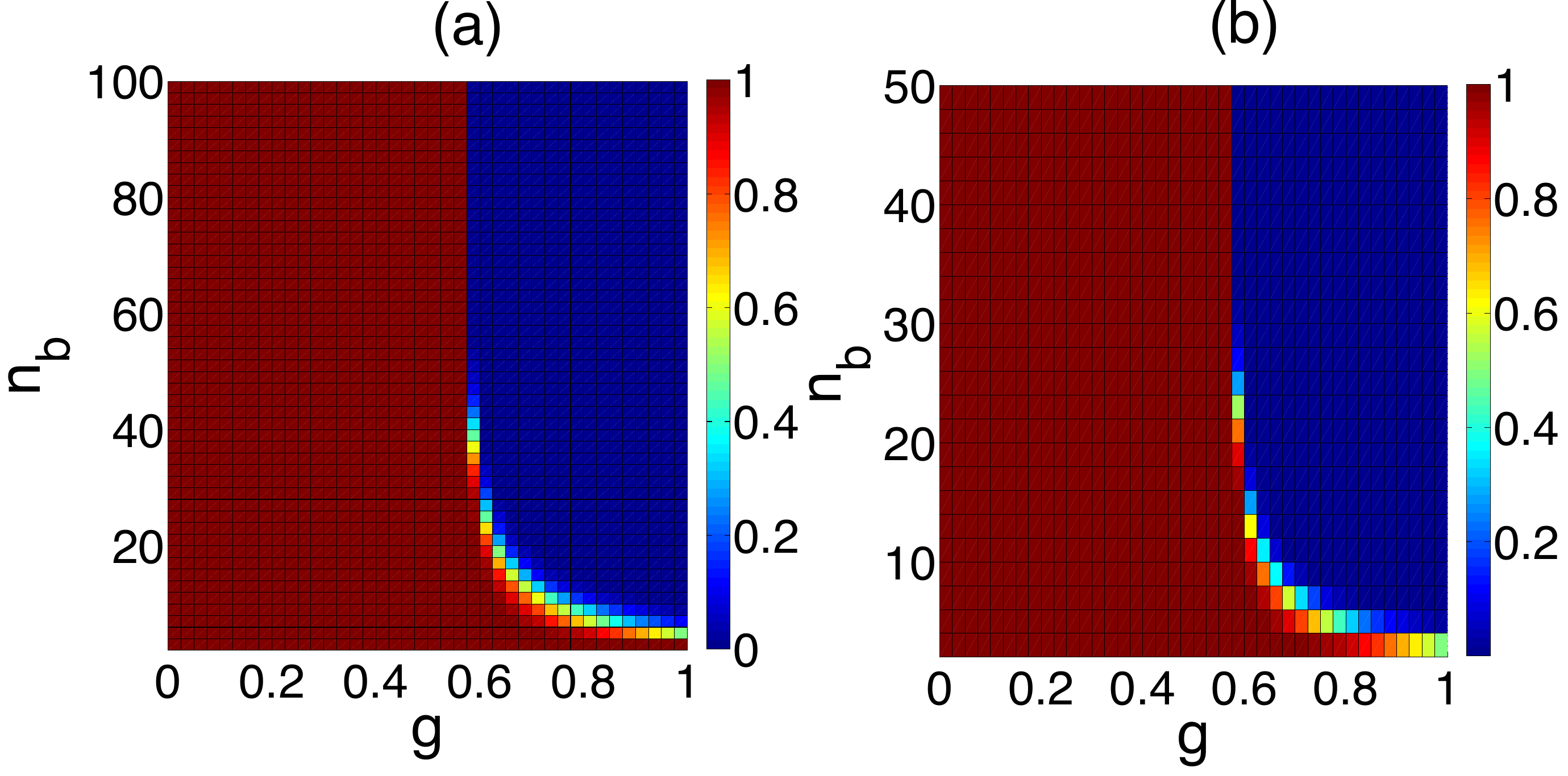}
  \caption{\label{AppCoarsed}
  (color online) $|E_\gamma|$ extrapolated from the scaling with $L$, as a function of the string tension $g$ and the number of cylinders $n_b$, for toruses up to $100 \times 100$ sites, for the perturbed toric code on a square lattice and cylinders of width (a) $l=1$, and (b) $l=2$. Notice that both plots look almost identical, but the vertical axis in (b) spans half the values of the vertical axis in (a).}
\end{figure}

As a proof of principle, we have done a calculation for cylinders with a width of more than one site, namely $l=2$. For toruses up to $100 \times 100$ sites, this implies that $n_b$ reaches up to $50$ cylinders. In Fig.\ref{AppCoarsed} we see an example of such a calculation for the perturbed toric code ground state on the square lattice, and compare it to that of Fig.\ref{Fig4}(a). As expected, we see convergence with $n_b$ twice as fast as compared to the case $l=1$, showing the same estimation for the quantum critical point $g^* \approx 0.56$.

\subsection{5. Alternative optimization strategies and finite-size effects}

The numerical optimization of tensors presented in the main text deals directly with systems of infinite size and then wraps them around a finite-size torus. Here we explore the alternative option of optimizing the tensors in the product state approximation to $\ket{\Psi(n,L)}$ directly on a torus, i.e., implementing the effect of periodic boundary conditions. We expect this optimization to be less efficient, therefore we can only access system sizes smaller than the ones mentioned in the main text. However, we also expect that finite-size corrections are more accurate, especially around the phase transition point. Thus, this is a valid and precise option to study finite-size corrections in $E_G$ for small- and medium-size systems.

\begin{center}
\begin{figure}
\includegraphics[width=12cm]{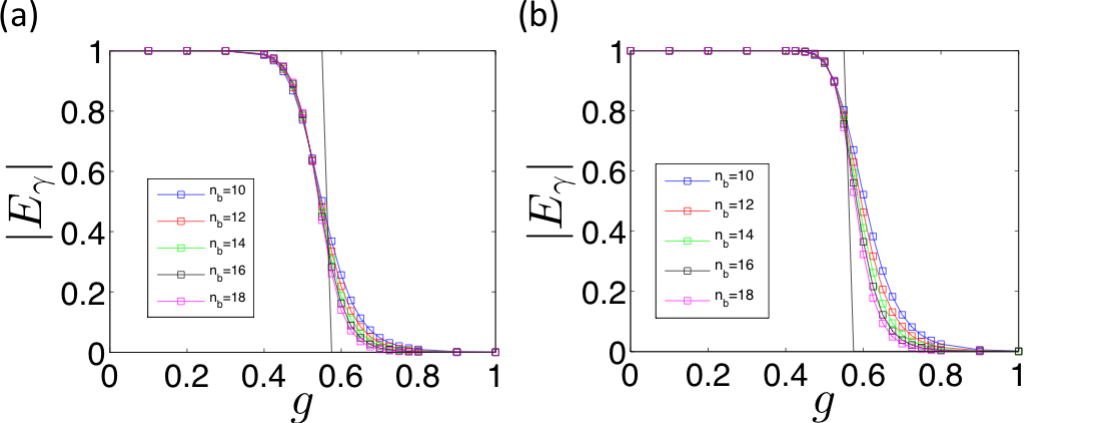}
\caption{(color online) $E_\gamma$ estimated from the finite-size optimization, for the square lattice toric code state $\ket{0,0}$, as a function of the string tension g and $n = 10 \ldots , 18$. (a) $L = 20$, and (b) $L = 30$. Notice that there is a strong dependence on $L$, especially around the transition region. As expected, larger systems show a more robust topological phase, in turn tending towards the results in the main text for $L \gg 1$, which are represented by a black solid line.}\label{sup:finite}
\end{figure}
\end{center}

Operating first with an exact representation of the ground state and of product states on each cylinder of the torus we observe high accuracy of the MPS representation. This calculation, thus, validates our MPS approximation for the states within each cylinder. For toruses up to $L\approx10$ and $n \gg 1$, we recover essentially the same results whenever the MPS bond dimension fulfills $\chi \approx 10-12$ (results not shown). Once this is clear, we have proceeded with the ``finite-size" optimization over MPSs for larger toruses. As one can see in Fig.\ref{sup:finite}, there is a strong dependence of $E_\gamma$ on the system size in the transition region, as expected. The transition appears at larger $g$ for increasing system size, while increasing the $n_b$ makes this transition sharper. This can be used as an extra justification for the method used in the main text, which hits very efficiently the infinite-size limiting behavior even around the transition region. In a nutshell, by using systems of large size we rule out the possibility of any finite-size effect coming either from the optimization of the tensors or from the finite-size scaling on the torus.

%To check the validity of this ``finite-size" optimization using MPSs for each cylinder, we have checked that we obtain essentially the same results as the exact optimization, where no MPS approximation is used for the cylinders. This calculation, thus, validates our MPS approximation for the states within each cylinder. For toruses up to $L\approx10$ and $n \gg 1$, we recover essentially the same results whenever the MPS bond dimension fulfills $\chi \approx 10-12$ (results not shown). Once this is clear, we have proceeded with the ``finite-size" optimization over MPSs for larger toruses. As one can see in Fig.\ref{sup:finite}, there is a strong dependence of $E_\gamma$ on the system size in the transition region, as expected. This can be used as an extra justification for the method used in the main text, which hits very efficiently the infinite-size limiting behavior even around the transition region. In a nutshell, by using systems of large size we rule out the possibility of any finite-size effect coming either from the optimization of the tensors or from the finite-size scaling on the torus.

\begin{center}
\begin{figure}
\includegraphics[width=7cm]{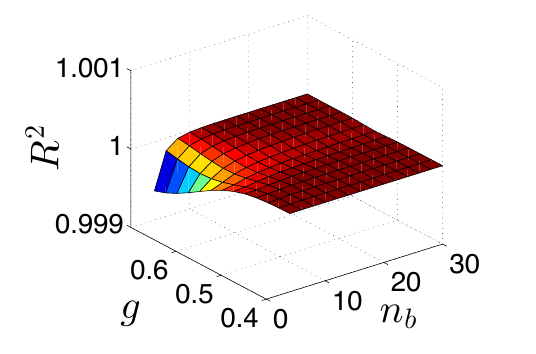}
\caption{(color online) $R^2$ for the linear fit. We observe that only at small system sizes the fit slightly deviates from almost-perfect linearity. The largest deviation is around the transition point, and vanishes for increasing system sizes.}\label{sup:l1}
\end{figure}
\end{center}

In this work we have computed $E_\gamma$ assuming a linear scaling with $L$ for $E_G$ and large $L$. Here we justify this
approach by showing that non-linear corrections vanish in the large-size limit. We test directly the validity of the linear fit by computing its coefficient of determination $R^2$. This is shown in Fig.\ref{sup:l1}, for systems up to $L=20$ and $n_b=30$. The plot shows $R^2$ as a function of $g$ and $n_{b}$, and and one can see that a significant deviation from a perfect linear fit (for which $R^2 = 1$) happens only close to the transition point, and in fact vanishes as the system size increases.

\begin{center}
\begin{figure}[h]
\includegraphics[width=12cm]{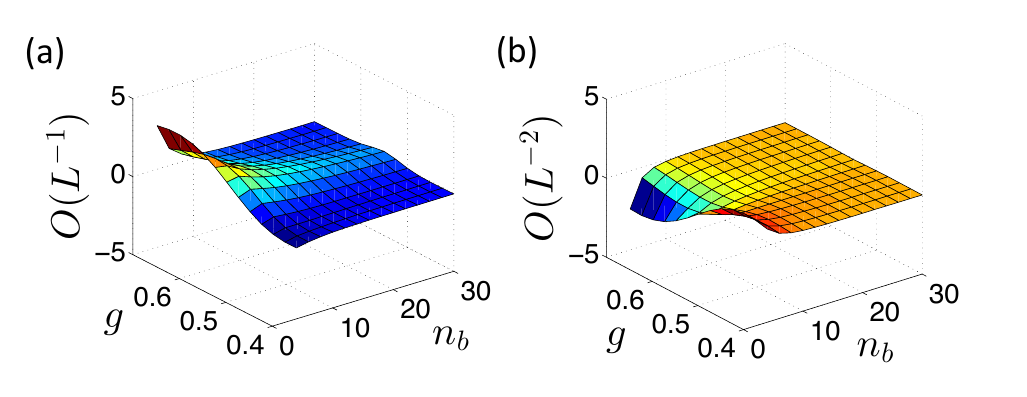}
\caption{(color online) (a) Contribution $\mathcal{O}(L^{-1})$ to the fit in Eq.\ref{sup:eq1}. Significant non-linear corrections appear only around the transition point, and vanish with for increasing sizes. (b) Contribution $\mathcal{O}(L^{-2})$ to the same fit. For $n_b\approx10$, the correction is already negligible even around the transition point. }\label{sup:l2}
\end{figure}
\end{center}

Finally, in order to assess the validity of the linear fit we can also study deviations from it, e.g.,
\begin{equation}\label{sup:eq1}
E_G = E_0 - E_\gamma + \mathcal{O}(L^{-1}) + \mathcal{O}(L^{-2}),
\end{equation}
which introduces non-linear terms in the fit. In Fig.\ref{sup:l2} we plot  the coefficients for the $\mathcal{O}(L^{-1})$ and $\mathcal{O}(L^{-2})$ contributions to Eq.\ref{sup:eq1}. We see that the presence of these two terms is only significant for small systems around the transition point, vanishing for large system sizes. Notice also that the results shown in the main text were obtained for large systems, so that these non-linear contributions should be negligible. Such finite-size effects are only accessible in small- and medium-size systems, as shown here.

\subsection{6. No continuous entanglement-loss in topological phases and local fidelity}

\begin{figure}
 \includegraphics[width=12cm]{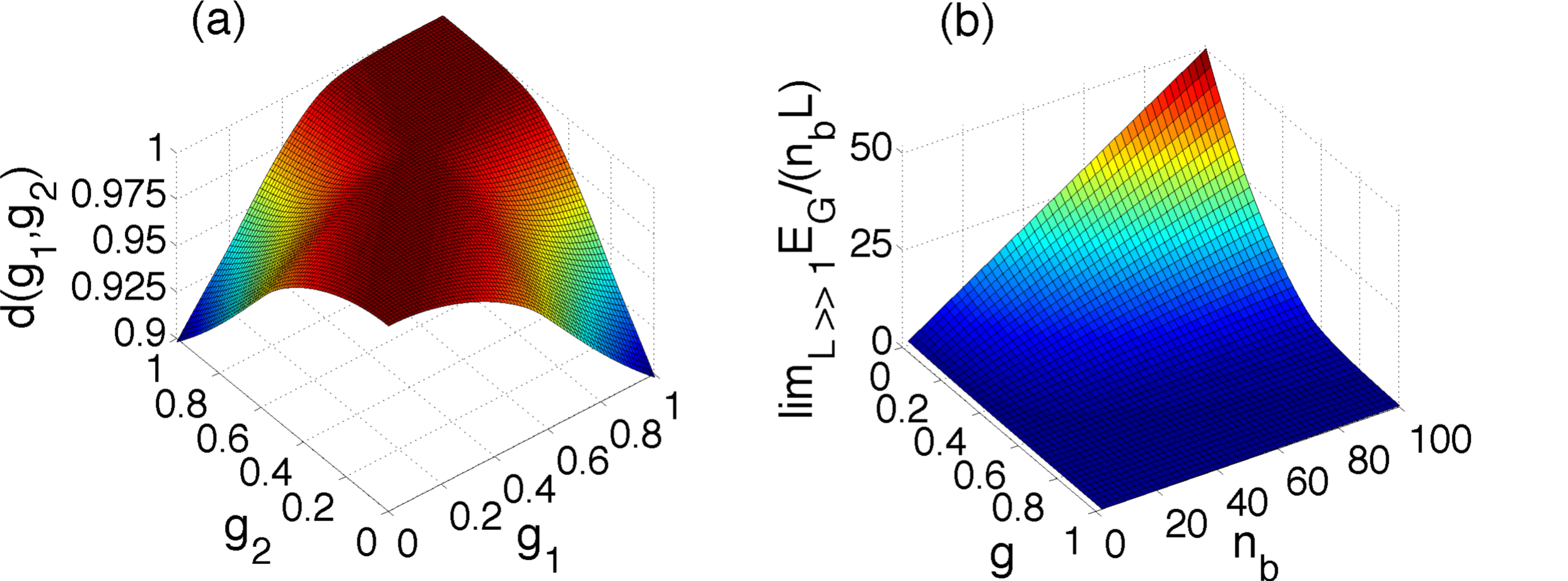}
  \caption{\label{Fig5}
  (color online) perturbed toric code on the square lattice: (a) local fidelity diagram. Similar results are obtained for the honeycomb lattice. (b) Coefficient of $L$ in the linear fit for large $L$, per block.}
\end{figure}

In Ref.\cite{aentloss} it was argued that 1d many-body systems with conformally-invariant quantum critical points display
continuous entanglement-loss along RG flows between fixed points. This was understood as
a refinement of Zamolodchikov's famous $c$-theorem \cite{acthe},
recently generalized to 2d \cite{a2dcthe}. However, recent works have
shown counter-examples to this behavior for different systems \cite{aloss}. Here we provide similar results using the density of GE per block, which for large $L$
corresponds to the linear term in our fits. To determine the nature of the RG flows in our systems, we computed the local fidelity diagram in Fig.\ref{Fig5}(a) \cite{afid}. The pinch-point in
the local fidelity $d(g_1,g_2)$ \cite{afidel} is in accordance with a continuous phase transition at $g^* \approx 0.56$, in turn corresponding to a critical and non-topological RG fixed point. Two more non-critical fixed points are present at $g=0$, which is topological, and $g = \infty$, which is trivial. Coarse-graining drives the quantum state towards either $g = 0$ or $g = \infty$ depending on the phase, and thus $g = g^*$ is an unstable fixed point. As for the density of GE per block, in Fig.\ref{Fig5}(b) we
see that, in the topological phase, it increases as we
move away from the phase transition point, so that the entanglement
per site rises when flowing from $g = g^*$ towards $g=0$. Albeit not a universal contribution, this already implies that  there is no continuous reordering of quantum correlations at
every infinitesimal RG step along the corresponding RG flow between these two fixed points, which rules out the possibility of continuous entanglement loss in the wave function.

\subsection{7. Correlation length critical exponent $\nu$ from $E_\gamma$}

Near criticality the only relevant length scale in the system is the correlation length, which scales as $\xi \sim |g-g^*|^{-\nu}$. For a finite-size system, the estimate $g_c$ of the critical point $g^*$ will depend on the system size $L_s$: $|g_c-g^*| \sim L_s^{-1/\nu}$. In our case, we have  two different sizes $n_b$ and $L$. If we take $L$ very large, the remaining finite length-scale will be given by $n_b$, and thus in our case $|g_c-g^*| \sim n_b^{-1/\nu}$. We can test such an ansatz scaling to estimate the critical exponent $\nu$ from the behavior of $E_\gamma (n_b,L\rightarrow \infty)$, i.e., from the data in Fig.\ref{Fig4}(a-c).

\begin{figure}
 \includegraphics[width=14cm]{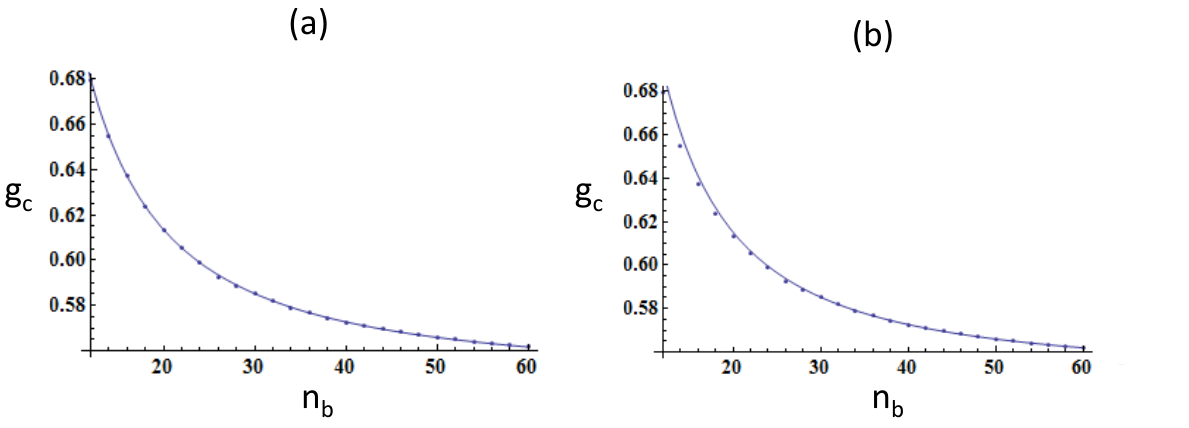}
  \caption{\label{AppCrit}
  (color online) (a) correlation length fit for $n_b$ between $12$ and $60$; (b) correlation length fit for $n_b$ between $40$ and $60$ (whole range of $n_b$ displayed).}
  \label{FigFit}
\end{figure}

For instance, consider the perturbed toric code on the square lattice (Fig.\ref{Fig4}(a)). For $n_b$ between $12$ and $60$ we find that the best fit is $g_c = 0.55 + 3.83/n_b^{1.34}$ with $R^2 = 0.99992$, thus $\nu\sim 1/1.34$ (Fig.\ref{FigFit}(a)). Between $n_b = 40$ and $60$, the best fit gives $g_c = 0.55 + 5.58/n_b^{1.48}$ with $R^2 = 0.99999$, giving $\nu\sim 1/1.48$, close to 2/3 (Fig.\ref{FigFit}(b)). The critical exponent $\nu$ is thus estimated to fall roughly between $0.68$ and $0.75$. For comparison, other values found in topological transitions for the square-lattice toric code model are the Ising value $\nu \sim 0.63$ for a parallel magnetic field, and also any value between $\nu \sim 0.63$ and $\nu \sim 1$ along a multicritical line for an arbitrary field \cite{expo}. In our case, we believe that errors come from the finite number of grid points for $g$ and the procedure of estimating $g_c$ (which we determine as the value of $g$ at $|E_{\gamma}| \sim 0.5$ by interpolating the available data). The accuracy can be improved by computing more data, but in any case, the analysis suggests that the transition is continuous, in accordance with our fidelity results in Fig.(\ref{Fig5}).

\subsection{8. Tensor network methods to compute R\'enyi entropies on infinite cylinders}

The R\'enyi entropy between a subsystem with reduced density matrix $\rho$ and the rest of the system is given by
\beq
S^{(n)} = \frac{1}{1-n}\log \left( \tr{(\rho^n)} \right),
\eeq
with $n$ the R\'enyi entropy index. The limit $n \rightarrow 1$ coincides with the usual von Neumann (or entanglement) entropy $S^{(1)} = -\tr{\rho \log \rho}$, whereas the limit $n \rightarrow \infty$ corresponds to the so-called infinity R\'enyi entropy, or single-copy entanglement, $S^{(\infty)} = - \log \nu_1$, with $\nu_1$ the largest eigenvalue of $\rho$. In Ref.\cite{aRenyi}Ê it was proven that all these entropies have the same topological contribution, so that any of them can be used to identify the topological nature of a given state.

Because of this, in combination with their apparent simplicity, R\'enyi entropies have become quite common tools to evaluate the presence of topological order in quantum many-body systems. For instance, the 2-R\'enyi entropy is a usual approach in Quantum Monte Carlo \cite{acost}, whereas in tensor networks, and especially in 2d PEPS, both the the 2-R\'enyi entropy and the infinite R\'enyi entropy (or single-copy entanglement) are easily accessible quantities. This follows, e.g., from the approach explained in Ref.\cite{aig}, which relies on wrapping the 2d PEPS around a cylinder of circumference $L$, and considering a non-contractible bipartition of the cylinder in two halves.

\begin{figure}
 \includegraphics[width=10cm]{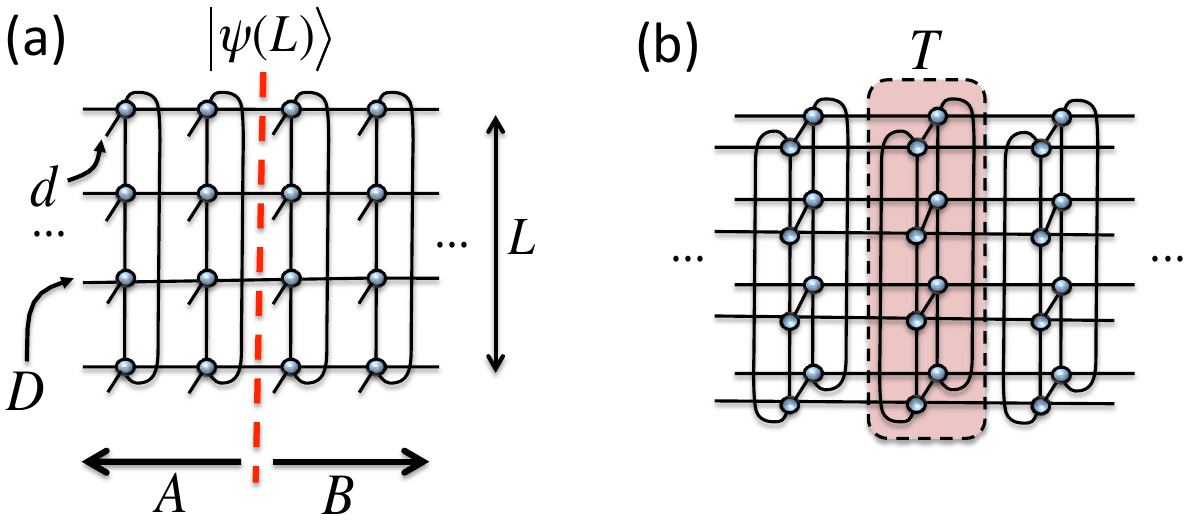}
  \caption{\label{AppCyl}
  (color online) (a) 2d PEPS on an infinite cylinder of circumference $L$. (b) Transfer matrix $T$.}
\end{figure}

Let us be more precise with the tensor network calculation: we start with a 2d PEPS $\ket{\psi(L)}$ on a cylinder of circumference $L$. For simplicity, we assume that the cylinder is infinitely long, though finite systems can also be considered easily. We partition the cylinder into two half-infinite pieces $A$ and $B$, see Fig.\ref{AppCyl}(a). As explained in Ref.\cite{aig}, the reduced density matrix of half an infinite cylinder (e.g. $A$) is given by
\beq
\rho = U \sqrt{\sigma_A^T}Ê\sigma_B \sqrt{\sigma_A^T} U^\dagger,
\eeq
with $\sigma_{A/B}$ the reduced density operators in $A/B$ for the virtual spaces across the bipartition, and $U$ an isometry. When the appropriate symmetries are present, as in the cases analyzed here, one has $\sigma_A = \sigma_B \equiv \sigma$, and therefore
\beq
\rho = U \sigma^2 U^\dagger.
\eeq
Notice that this readily implies that $\rho$ and $\sigma^2$ are isospectral, since the isometries do not change non-zero eigenvalues, and therefore the R\'enyi entropies only depend on the eigenvalues of $\sigma$. And what is more: in terms of $\sigma$ the R\'enyi entropies now read
\beq
S^{(n)} = \frac{1}{1-n}\log \left( \tr{(\sigma^{2n})} \right).
\label{sig}
\eeq
Thus, computing $S^{(n)}$ amounts to calculating $\sigma$, its powers, and its trace \footnote{In fact, with this approach one could also compute $S^{(1/2)}, S^{(3/2)}, ...$.}.

The calculation of $\sigma$ on an infinite cylinder for a 2d PEPS is a well-posed tensor network problem, since $\sigma$ is nothing but the dominant (left or right) eigenvector of the PEPS transfer matrix $T$, see Fig.\ref{AppCyl}(b). This can be solved using many different strategies. Here we choose a similar approach to the one described in the main text for the GE: we compute this dominant eigenvector using the iTEBD method for non-unitary evolutions \cite{aitebd1, aitebd2}. The resulting dominant eigenvector can be written as an MPO of bond dimension $\chi$, which is then wrapped around a circle of length $L$, and constitutes our approximated $\sigma$. The computational cost of this calculation is $O(\chi^3 D^6 + \chi^2 D^8)$ \cite{aiPEPS}.

Next, by combining the obtained MPO for $\sigma$ with Eq.\ref{sig} it is easy to see that, again, the calculation of any R\'enyi entropy $S^{(n)}$ can be reduced to some standard tensor network problem, see Figs.\ref{AppSig} and \ref{AppSig2}. From here, the calculation strategy differs depending on whether we are interested in finite-$n$, or in the limit $n \rightarrow \infty$. The details of both cases are as follows:

\subsubsection{Finite-$n$ R\'enyi entropies}

\begin{figure}[h]
 \includegraphics[width=11cm]{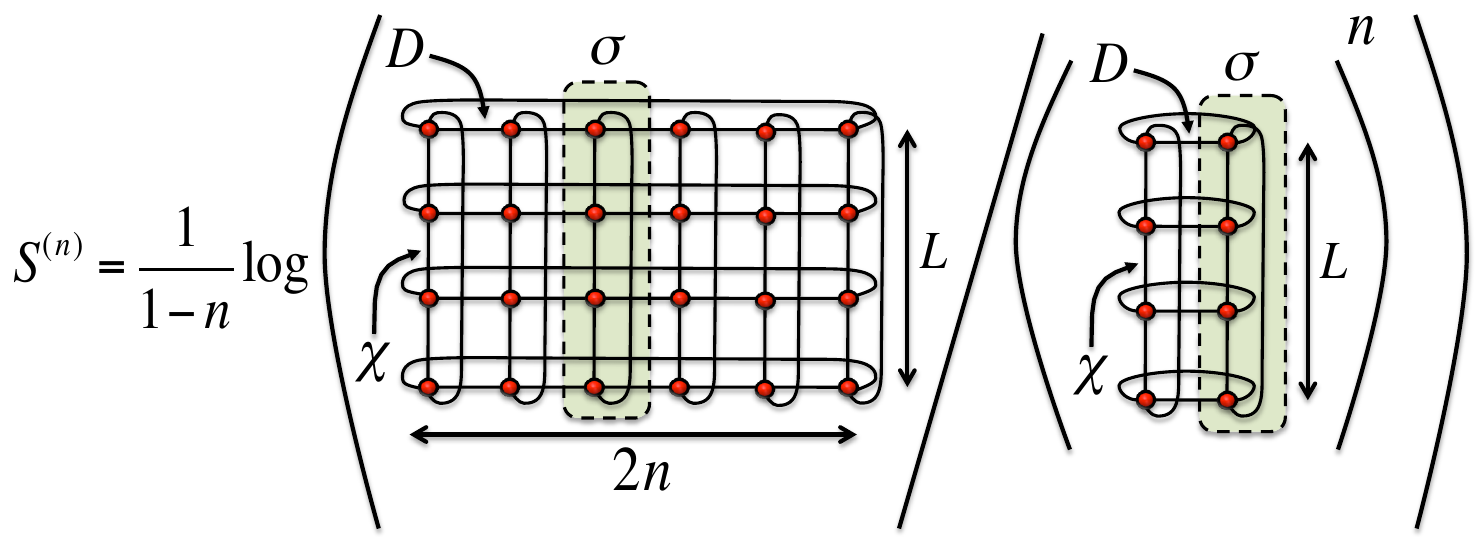}
  \caption{\label{AppSig}
  (color online) tensor network diagrams corresponding to the finite-$n$ R\'enyi entropy.}
\end{figure}

In this case the calculation amounts to computing the tensor network contraction from Fig.\ref{AppSig}, taking into account the normalization of $\sigma^2$. Formally, the numerator inside the logarithm amounts to the contraction of a 2d tensor network on a torus of size $n \times L$, thus being quite similar to some of the contractions that were involved  in the calculation of the GE in the main text. Therefore, we simply use now the same approach that was used there, i.e., we approximate the repeated multiplication of MPOs using iTEBD as explained in the first section of this supplementary material. After exactly $2n$ iterations we stop, and wrap the resulting MPO with bond dimension $\chi'$ around a circle of size $L$. The total computation time for this approach scales as $O(2n(\chi^2 \chi'^2 D^3 + \chi^3 \chi'^3 D^2) + L\chi'^3)$.

\subsubsection{$\infty$-R\'enyi entropy (single-copy entanglement)}

\begin{figure}[h]
 \includegraphics[width=10cm]{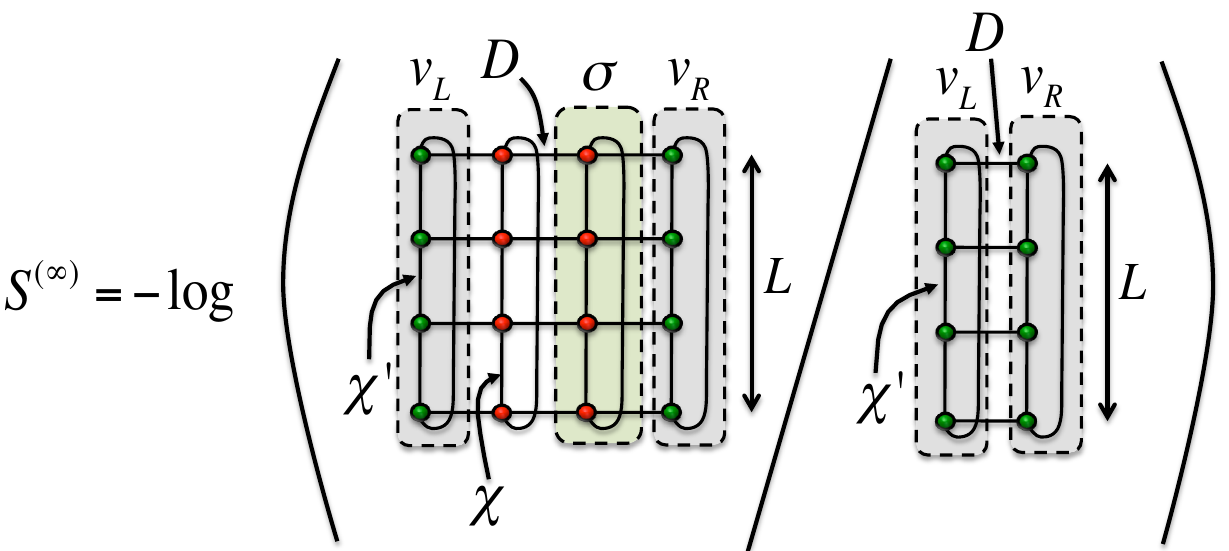}
  \caption{\label{AppSig2}
  (color online) tensor network diagrams corresponding to the $n=\infty$ R\'enyi entropy.}
\end{figure}

The case $n = \infty$ deserves special attention. In principle, it should be possible to use the same trick as for finite-$n$ above, taking the limit of a very large $n$ until some convergence is reached. However, even if correct, such an approach is not necessarily the most efficient one. To understand this, simply notice that the largest eigenvalue $\nu_1$ of $\rho$ also corresponds to the largest (normalized) eigenvalue of $\sigma^2$, i.e.,
\beq
\nu_1 = \frac{v_L^T  \sigma^2  v_R}{v_L^T v_R},
\eeq
with $v_L$ and $v_R$ respectively the left and right dominant eigenvectors of $\sigma$. Since $\sigma$ is given in the form of an MPO with periodic boundary conditions, it turns out that finding such dominant eigenvectors is also a standard 2d tensor network problem which we solve as done many times before: using iTEBD and wrapping the resulting MPS for $v_L$ and $v_R$, with bond dimension $\chi'$, around a circle of size $L$. In the end, $S^{(\infty)}$ is computed as in the diagram in Fig.\ref{AppSig2}. The overall computational cost of this approach is $O(\chi^3 \chi'^3 D^3 + L \chi^5 \chi'^5 D)$.

\subsubsection{Remarks}
Some remarks are in order. First, if we are interested in extracting the topological component, we just proceed by doing finite-size scaling with $L$, fitting the results to a linear function, and extrapolating the result down to $ L \rightarrow 0$, exactly as we did for the GE. Second, the calculations of R\'enyi entropies with 2d PEPS can of course be improved in accuracy, especially if the correlation length is large (e.g. close to criticality). However, the practical drop in efficiency is very significant. And what is more: the same improvements that can be applied to R\'enyi entropy calculations can also be used, if necessary, to improve the accuracy in computing the GE. Third, we have observed that the calculations of the $n$-th R\'enyi entropies are extremely sensitive to truncations in the associated bond dimensions, which becomes especially important when $n$ gets `large". In practice, this means that we have been unable to produce sensible results for the topological contribution of R\'enyi entropies for $n > 3$ (except for infinity) because we cannot reach sufficiently large bond dimensions with our computing resources. This is to be contrasted with the remarkable robustness and accuracy of the GE, already for small bond dimensions. In other words: for a comparable value of bond dimensions and computation time, the topological GE outperforms in accuracy the topological R\'enyi entropies by far, especially close to a quantum critical point.

\color{black}
\subsection{9. Truncation errors in $E_G$ and R\'enyi entropies.}

\begin{figure}
 \includegraphics[width=12cm]{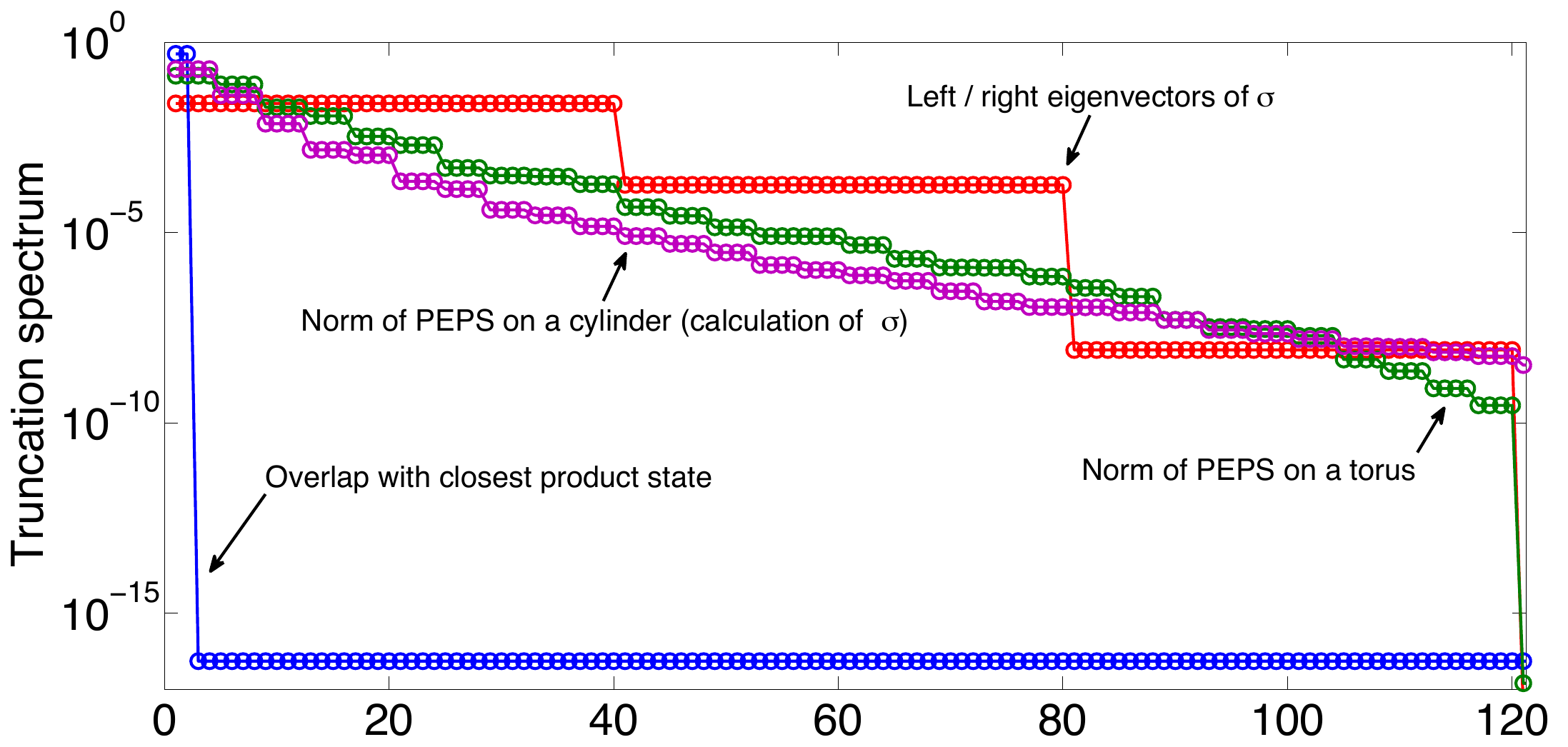}
  \caption{\label{figextra2}
  (color online) Different spectra to be truncated in the evaluation of $E_G$ or the R\'enyi entropies, for the perturbed  toric code on a square lattice with $g=0.55$.}
\end{figure}

As explained in the main text, our method to compute $E_G$ uses mainly single-layer contractions of a $2d$ tensor network. Such contractions appear in the calculation of the overlap between the optimal product state and the $2d$ PEPS, and can be computed very efficiently and as well as very accurately. Moreover, they are also numerically more  stable than the usual double-layer contractions needed to compute, e.g., the norm of the $2d$ PEPS. Almost half of the numerical manipulations to evaluate $E_G$ rely on such single-layer contractions. Moreover, the necessary double-layer contractions to evaluate $E_G$ are only those to compute the norm of a PEPS, which is a single scalar quantity (unlike an operator that is re-used). All in all, this makes the evaluation of $E_G$ inherently very accurate and robust. A different scenario is present for R\'enyi entropies, where the only fundamental object $\sigma$ is computed from a double-layer contraction. Such object $\sigma$ is further used in more calculations, which in practice means that there is a concatenation of errors if $\sigma$ itself is not computed extremely accurately. In practice, this means that R\'enyi entropies can be prone to a larger error than the $E_G$ whenever we have quantum states with a lot of entanglement, especially near a quantum phase transition, and this is indeed what we see in the main text.  

To further justify this picture, we have plotted in Fig.\ref{figextra2} the first $120$ values of the spectrum to be truncated in different steps of the calculations mentioned above. In the figure one can see that, while the spectrum for the single-layer contraction decays extremely fast, all the rest decay very slowly. As claimed, this means that truncation effects are less dramatic for $E_G$, which depends partly on truncations of single-layer calculations spectra and do not re-use objects computed from double-layer calculations. Since such single-layer spectra decay very fast, they can also be truncated accurately with a small truncation parameter, and hence the overall calculation is nor only more accurate, but also faster than the one of any of the R\'enyi entropies, as claimed. 

\color{black}


\begin{thebibliography}{2}
\bibitem{afidel}
If $\ket{\Psi(g)}$ is the PEPS with parameter $g$ on a 2d lattice of $N$ sites, then $|\braket{\Psi(g_1)}{\Psi(g_2)}| \sim d(g_1,g_2)^N$ for $N \gg 1$.
\end{thebibliography}
\end{document}